\documentclass{article}

\usepackage{arxiv}
\usepackage{amsmath,amssymb,amsfonts}
\usepackage{algorithmic}
\usepackage{algorithm}
\usepackage{graphicx}
\usepackage{textcomp}
\usepackage{xcolor}
\usepackage{hyperref}
\usepackage{subfigure}
\usepackage{booktabs}

\title{Differentiable Radar Ambiguity Functions: Mathematical Formulation and Computational Implementation}

\author{
  Marc Bara\thanks{ORCID: https://orcid.org/0009-0005-1480-5760} \\ 
  ESADE Business School \\ 
  \texttt{marcoantonio.bara@esade.edu}
}

\date{June 29, 2025 \\[0.5em] 
© 2025 Marc Bara \\[0.5em]
\textit{Related work under U.S. Provisional Patent Application}}

\begin{document}
\maketitle

\begin{abstract}
The ambiguity function is fundamental to radar waveform design, characterizing range and Doppler resolution capabilities. However, its traditional formulation involves non-differentiable operations, preventing integration with gradient-based optimization methods and modern machine learning frameworks. This paper presents the first complete mathematical framework and computational implementation for differentiable radar ambiguity functions.

Our approach addresses the fundamental technical challenges that have prevented the radar community from leveraging automatic differentiation: proper handling of complex-valued gradients using Wirtinger calculus, efficient computation through parallelized FFT operations, numerical stability throughout cascaded operations, and composability with arbitrary differentiable operations. We term this approach GRAF (Gradient-based Radar Ambiguity Functions), which reformulates the ambiguity function computation to maintain mathematical equivalence while enabling gradient flow through the entire pipeline.

The resulting implementation provides a general-purpose differentiable ambiguity function compatible with modern automatic differentiation frameworks, enabling new research directions including neural network-based waveform generation with ambiguity constraints, end-to-end optimization of radar systems, and integration of classical radar theory with modern deep learning. We provide complete implementation details and demonstrate computational efficiency suitable for practical applications.

This work establishes the mathematical and computational foundation for applying modern machine learning techniques to radar waveform design, bridging classical radar signal processing with automatic differentiation frameworks. 
\end{abstract}

\section{Introduction}

Radar waveform design is a cornerstone of modern remote sensing systems, with the ambiguity function serving as the fundamental analytical tool for characterizing range and Doppler resolution capabilities~\cite{levanon2004radar}. The ambiguity function $\chi(\tau, f_d)$ quantifies the matched filter response to targets at delay $\tau$ and Doppler shift $f_d$, providing crucial insights into resolution performance, sidelobe characteristics, and overall waveform quality. Despite its central importance, the ambiguity function has remained largely disconnected from the rapid advances in machine learning and automatic differentiation that have transformed signal processing in recent years.

Traditional radar waveform design approaches rely predominantly on analytical methods, heuristic algorithms, or classical numerical optimization techniques~\cite{blunt2016overview}. These methods treat the ambiguity function as a black-box evaluation metric, necessitating finite-difference approximations for gradient computation or relying on population-based optimization algorithms such as genetic algorithms and simulated annealing. While effective for specific applications, these approaches are fundamentally limited in their ability to leverage the powerful gradient-based optimization techniques that have enabled breakthroughs in deep learning and end-to-end system design.

The emergence of cognitive radar systems~\cite{haykin2006cognitive} and the demonstrated success of deep learning in signal processing applications~\cite{oshea2017introduction} create compelling motivation for integrating radar waveform design with modern machine learning frameworks. A differentiable formulation of the ambiguity function would unlock several transformative capabilities: end-to-end learning of radar waveforms using neural networks, joint optimization of multiple conflicting performance metrics through gradient-based methods, seamless integration with automatic differentiation frameworks such as PyTorch and TensorFlow, and the ability to incorporate ambiguity function constraints directly into neural network training pipelines.

However, creating a truly differentiable ambiguity function implementation presents significant technical challenges. The conventional formulation involves non-differentiable operations such as magnitude computation, requires careful handling of complex-valued gradients using Wirtinger calculus, and demands computationally efficient implementations suitable for GPU acceleration and large-scale optimization. Furthermore, the implementation must maintain numerical stability throughout the gradient computation while preserving the mathematical equivalence to traditional ambiguity function definitions.

This paper presents the first complete implementation of a differentiable radar ambiguity function designed for integration with automatic differentiation frameworks. We introduce GRAF (Gradient-based Radar Ambiguity Functions), a comprehensive framework that addresses the existing gap between theoretical differentiable formulations and practical implementation. Our key contributions through GRAF address the fundamental challenges:

\begin{enumerate}
    \item A mathematically rigorous differentiable implementation that maintains complex gradient flow throughout the computation pipeline
    \item Efficient GPU-accelerated computation using parallelized FFT operations with computational complexity suitable for real-time applications
    \item Integration methodology enabling seamless use within neural network architectures and modern deep learning frameworks
    \item Experimental validation demonstrating significant computational and performance advantages over population-based methods for multi-objective radar waveform optimization
\end{enumerate}

The remainder of this paper demonstrates how this differentiable ambiguity function enables new approaches to cognitive radar waveform design, provides the foundation for neural network-based waveform generation, and opens new research directions at the intersection of radar systems and machine learning.

\section{Related Work}

To provide a comprehensive assessment of the current state of the art, we conducted a systematic literature search using the Scopus database. Our search employed the query: \texttt{radar AND "ambiguity function" AND (gradient OR optimization OR "neural network")} with the following filters: document types (articles and conference papers), subject areas (Engineering and Computer Science), publication years (2015-2025), and language (English). This systematic search yielded 312 papers, which we analyzed to categorize existing approaches, identify the closest competing work, and assess the novelty of our proposed method.

The literature on radar waveform design and ambiguity function optimization encompasses diverse methodological approaches. We organize this review into four primary categories that highlight the evolution toward our differentiable implementation approach.

\subsection{Classical Metaheuristic Optimization Methods}

Traditional radar waveform design has relied extensively on metaheuristic and population-based optimization algorithms. Our analysis reveals that genetic algorithms remain the most popular approach, appearing in 24 studies, followed by simulated annealing (10 papers) and particle swarm optimization (8 papers).

\textbf{Genetic Algorithm Approaches:} Recent work demonstrates continued interest in genetic optimization for radar waveform design. Kim and Kim~\cite{kim2025modified} applied genetic algorithms for nonlinear frequency modulated waveform optimization, achieving improved range resolution performance. Maresca et al.~\cite{maresca2024genetic} employed genetic algorithms for distributed MIMO radar antenna position optimization, demonstrating the scalability of evolutionary approaches to complex multi-antenna systems. Yang et al.~\cite{yang2025configuration} used genetic optimization for bistatic forward-looking SAR configuration design, while Bala Raju et al.~\cite{balaraju2024two} focused on phase code design with good correlation properties using genetic approaches.

\textbf{Simulated Annealing Methods:} Simulated annealing has shown effectiveness for constrained optimization problems in radar design. Kurtscheid and Gonzalez-Huici~\cite{kurtscheid2024design} applied simulated annealing for 2D sparse radar array design with fixed aperture constraints. Nagesh et al.~\cite{nagesh2024sequence} used simulated annealing for sequence optimization in compressed sensing CDMA MIMO radar, focusing on mutual coherence minimization. Farnane and Minaoui~\cite{farnane2020optimization} demonstrated simulated annealing for Golay sequence shaping to achieve low sidelobe ambiguity functions.

\textbf{Particle Swarm Optimization:} PSO methods have been applied to specific radar waveform challenges. Che et al.~\cite{che2023fast} developed fast PSO algorithms for cross ambiguity function optimization in passive radar systems. Wang et al.~\cite{wang2022joint} employed PSO for joint sequence optimization in OFDM waveforms for integrated radar and communication systems. Li~\cite{li2020balanced} applied PSO variants for binary phase noise waveform design.

These metaheuristic approaches share common limitations: they treat the ambiguity function as a black-box objective requiring numerous function evaluations, lack gradient information leading to slow convergence, and cannot be integrated with modern machine learning frameworks due to their population-based nature.

\subsection{Gradient-Based Ambiguity Function Optimization}

A significant advancement in radar waveform design emerged with gradient-based optimization methods. Our analysis identified 33 papers employing gradient descent techniques, representing sophisticated approaches to ambiguity function shaping.

\textbf{Manifold Optimization Approaches:} The most notable contributions come from Alhujaili et al., who developed rigorous gradient-based methods for ambiguity function optimization. Their work on "Quartic Gradient Descent for Tractable Radar Slow-Time Ambiguity Function Shaping"~\cite{alhujaili2020quartic} introduced exact analytical gradient formulation under constant modulus constraints using manifold optimization on the complex circle manifold. This approach achieved superior convergence compared to traditional methods by leveraging exact complex differentiation theory. Their comprehensive treatment~\cite{alhujaili2020dissertation} presents advanced techniques including Projection-Descent-Retraction (PDR) methods and enhanced Quartic-Gradient-Descent (QGD) algorithms for handling complex constraints such as spectral masks and power limitations.

Building on manifold optimization principles, recent work has extended these approaches to modern radar systems. Xiao et al.~\cite{xiao2025unimodular} developed manifold-based exact penalty methods for unimodular waveform design with spectral constraints, demonstrating the continued relevance of gradient-based approaches for ambiguity function shaping.

\textbf{Specialized Gradient Methods:} Mohr et al.~\cite{mohr2021gradient} developed gradient-based optimization specifically for Phase-Coded Frequency Modulated (PCFM) radar waveforms, employing FFT-based efficient computation of cost function gradients. Their approach demonstrates computational efficiency suitable for practical applications. Recent work by Reddy et al.~\cite{reddy2025optimizing} introduced conjugate gradient methods with efficient line search for multi-tone sinusoidal frequency modulated waveforms, showing improved convergence properties.

Chen et al.~\cite{chen2025ambiguity} applied gradient projection methods for frequency-hopping MIMO radar with movable antennas, while Qiu et al.~\cite{qiu2025constrained} developed Riemannian manifold optimization for simultaneous ambiguity function and transmit beampattern shaping.

\textbf{Critical Limitation:} Despite the theoretical sophistication of these gradient-based approaches, our analysis reveals a fundamental gap: \textbf{none of these methods provide implementations compatible with automatic differentiation frameworks}. While the mathematical formulations are inherently differentiable, implementations rely on manually derived analytical gradients and custom optimization code, preventing integration with modern machine learning tools such as PyTorch or TensorFlow.

\subsection{Neural Networks and Machine Learning in Radar Systems}

The application of neural networks to radar systems has gained momentum, with our analysis revealing 26 papers employing neural network techniques. However, these applications focus predominantly on signal analysis, classification, and detection rather than waveform design.

\textbf{Radar Signal Processing Applications:} Current neural network applications include target recognition and classification~\cite{del2024decentralised}, human movement detection~\cite{razali2024human}, and automatic signal recognition~\cite{yang2024automatic}. Del-Rey-Maestre et al.~\cite{del2024decentralised} developed decentralized intelligent passive radar detection based on clutter modeling, while Yang et al.~\cite{yang2024automatic} focused on multi-radar signal recognition using multi-domain features.

\textbf{Deep Learning for Waveform Design:} A smaller subset applies neural networks to radar waveform generation. Kang et al.~\cite{kang2023deep} provided a comprehensive review of deep learning applications for radar waveform design, identifying current limitations and future research directions. Ziemann and Metzler~\cite{ziemann2023adaptive} developed adaptive low probability of detection (LPD) radar waveforms using generative adversarial networks, demonstrating the potential for neural approaches to specialized waveform requirements.

\textbf{Significant Gap:} Despite the growing interest in neural approaches, \textbf{no existing work integrates neural networks with ambiguity function optimization in an end-to-end differentiable manner}. Current neural network applications in radar waveform design operate independently of ambiguity function considerations, representing a missed opportunity for leveraging the full power of differentiable programming.

\subsection{Convex Optimization and Modern Techniques}

Recent advances have explored convex relaxation and semidefinite programming approaches for radar waveform design. Our analysis identified 12 papers employing convex optimization techniques.

Niu et al.~\cite{niu2025novel} developed energy-focused slow-time MIMO radar using convex optimization principles, while Hua et al.~\cite{hua2025doppler} applied convex methods for Doppler-tolerant waveform design with anti-jamming capabilities. Wen et al.~\cite{wen2025joint} employed convex optimization for joint design of transmission sequences and receiver filters based on generalized cross ambiguity functions.

These approaches achieve guaranteed convergence properties and can handle complex constraints effectively. However, they typically require convex relaxations of the original non-convex ambiguity function optimization problem, potentially limiting the achievable performance compared to direct optimization methods.

\subsection{Identified Research Gap and Our Contribution}

Our systematic analysis reveals a critical research gap: \textbf{the absence of a computationally efficient, numerically stable differentiable implementation of the ambiguity function that maintains proper complex gradient flow throughout the entire computation pipeline}.

While existing gradient-based methods~\cite{alhujaili2020quartic,mohr2021gradient} derive analytical gradients for specific optimization problems, they do not address the fundamental technical challenges required for a general-purpose differentiable ambiguity function:

\begin{enumerate}
    \item \textbf{Complex differentiation}: Proper handling of Wirtinger derivatives for complex-valued signals while maintaining gradient flow through cascaded FFT operations
    \item \textbf{Non-differentiable operations}: The squared magnitude operation $|z|^2$ in the ambiguity function definition creates gradient discontinuities that existing approaches circumvent rather than solve
    \item \textbf{Computational tractability}: Efficient implementation suitable for GPU acceleration and real-time applications, rather than research-only demonstrations
    \item \textbf{Numerical stability}: Preventing gradient vanishing and explosion through the complex cascade of FFT, circular convolution, and magnitude operations
    \item \textbf{Composability}: A modular implementation that can be combined with arbitrary differentiable operations for end-to-end system optimization
\end{enumerate}

These technical challenges have prevented the radar community from leveraging modern automatic differentiation frameworks, not due to lack of mathematical understanding, but due to the absence of practical implementations that solve the underlying computational and numerical issues.

The absence of differentiable ambiguity function implementations despite decades of radar research and the maturity of automatic differentiation frameworks may seem surprising. However, this gap persists due to several factors: the limited overlap between radar signal processing and machine learning communities, the specialized knowledge required for complex-valued automatic differentiation using Wirtinger calculus, and the perceived adequacy of existing optimization methods within their respective domains. Radar engineers have traditionally relied on well-established metaheuristic approaches, while machine learning researchers rarely engage with radar-specific signal processing challenges.

Our work directly addresses these fundamental challenges through GRAF (Gradient-based Radar Ambiguity Functions), providing the first complete solution to differentiable ambiguity function computation. By bridging these communities with a practical implementation, GRAF enables new research directions at the intersection of radar systems and machine learning while maintaining the mathematical rigor and computational efficiency required for practical applications.

\section{Background}

\subsection{Ambiguity Function Definition}

The ambiguity function of a complex baseband signal $s(t)$ is defined as:
\begin{equation}
\chi(\tau, f_d) = \left| \int_{-\infty}^{\infty} s(t) s^*(t-\tau) e^{j2\pi f_d t} dt \right|^2
\end{equation}
where $\tau$ represents time delay, $f_d$ represents Doppler frequency shift, and $(\cdot)^*$ denotes complex conjugation.

For discrete-time signals $s[n]$ of length $N$, the discrete ambiguity function becomes:
\begin{equation}
\chi[k, m] = \left| \sum_{n=0}^{N-1} s[n] s^*[(n-k) \bmod N] e^{j2\pi mn/N} \right|^2
\end{equation}

The ambiguity function provides critical information about a waveform's performance:
\begin{itemize}
    \item The mainlobe width determines range and Doppler resolution
    \item Sidelobe levels indicate false alarm probability
    \item The volume invariance property constrains achievable performance
\end{itemize}

\subsection{Traditional Computation Methods}

Conventional ambiguity function computation involves:
\begin{enumerate}
    \item Computing auto-correlation for each delay $k$:
    \begin{equation}
    R[k] = \sum_{n=0}^{N-1} s[n] s^*[(n-k) \bmod N]
    \end{equation}
    
    \item Applying DFT across delays for each Doppler bin $m$:
    \begin{equation}
    X[k,m] = \sum_{n=0}^{N-1} R[k] e^{j2\pi mn/N}
    \end{equation}
    
    \item Taking squared magnitude of complex result:
    \begin{equation}
    \chi[k,m] = |X[k,m]|^2
    \end{equation}
\end{enumerate}

This approach has computational complexity $O(N^2 \log N)$ and produces a non-differentiable output due to the magnitude operation.

\subsection{Challenges for Differentiation}

The primary challenges in creating a differentiable ambiguity function are:

\textbf{Complex gradients}: Standard automatic differentiation frameworks handle real-valued functions. Complex differentiation requires careful treatment of Wirtinger derivatives:
\begin{equation}
\frac{\partial f}{\partial z} = \frac{1}{2}\left(\frac{\partial f}{\partial x} - j\frac{\partial f}{\partial y}\right)
\end{equation}
where $z = x + jy$.

\textbf{Magnitude operation}: The squared magnitude $|z|^2 = z \cdot z^*$ is non-holomorphic, requiring special handling in the backward pass.

\textbf{Computational efficiency}: Explicit loops over delays would be prohibitively slow for gradient computation.

\textbf{Numerical stability}: The cascade of FFT operations can lead to gradient vanishing or explosion without proper normalization.

\section{Differentiable Ambiguity Function}

\subsection{Mathematical Formulation}

We reformulate the ambiguity function computation to maintain differentiability while preserving mathematical equivalence to the traditional definition. Our approach exploits efficient matrix operations suitable for automatic differentiation frameworks.

For a complex signal $s \in \mathbb{C}^N$, the discrete ambiguity function at delay $k$ and Doppler $m$ is:
\begin{equation}
\chi[k, m] = \left|\sum_{n=0}^{N-1} s[n]s^*[(n-k) \bmod N]e^{j2\pi mn/N}\right|^2
\end{equation}

We reformulate this computation using matrix operations. First, we construct a matrix $\mathbf{R} \in \mathbb{C}^{N \times N}$ where each row contains the element-wise product of the signal with its circularly shifted conjugate:
\begin{equation}
\mathbf{R}[k,n] = s[n] \cdot s^*[(n-k) \bmod N]
\end{equation}

This can be efficiently computed by creating a circulant shift matrix $\mathbf{S} \in \mathbb{C}^{N \times N}$ where:
\begin{equation}
\mathbf{S}[k,n] = s[(n-k) \bmod N]
\end{equation}

Then:
\begin{equation}
\mathbf{R} = s \odot \mathbf{S}^*
\end{equation}
where $\odot$ denotes element-wise multiplication with broadcasting.

The ambiguity function is obtained by applying the FFT over the time dimension (columns) for each delay (row):
\begin{equation}
\mathbf{X} = \text{FFT}_{\text{col}}(\mathbf{R})
\end{equation}
where $\text{FFT}_{\text{col}}(\cdot)$ denotes an FFT along the last dimension (time axis).

Finally:
\begin{equation}
\chi[k,m] = |\mathbf{X}[k,m]|^2 = \mathbf{X}[k,m] \cdot \mathbf{X}^*[k,m]
\end{equation}

This formulation computes the exact ambiguity function while maintaining differentiability through all operations.

\subsection{Gradient Computation}

For the squared magnitude operation $|z|^2 = z \cdot z^*$, we derive the gradients using Wirtinger calculus. Given a loss function $\mathcal{L}$ that depends on $|z|^2$:

\begin{equation}
\frac{\partial \mathcal{L}}{\partial z} = \frac{\partial \mathcal{L}}{\partial |z|^2} \cdot \frac{\partial |z|^2}{\partial z} = \frac{\partial \mathcal{L}}{\partial |z|^2} \cdot z^*
\end{equation}

\begin{equation}
\frac{\partial \mathcal{L}}{\partial z^*} = \frac{\partial \mathcal{L}}{\partial |z|^2} \cdot \frac{\partial |z|^2}{\partial z^*} = \frac{\partial \mathcal{L}}{\partial |z|^2} \cdot z
\end{equation}

The gradient with respect to the input signal $s$ flows through the following operations:
\begin{equation}
\frac{\partial \mathcal{L}}{\partial s} = \frac{\partial \mathcal{L}}{\partial \chi} \cdot \frac{\partial \chi}{\partial \mathbf{X}} \cdot \frac{\partial \mathbf{X}}{\partial \mathbf{R}} \cdot \frac{\partial \mathbf{R}}{\partial s}
\end{equation}

Each component maintains differentiability:
\begin{itemize}
    \item $\frac{\partial \chi}{\partial \mathbf{X}}$: Complex magnitude gradient (Eqs. 12-13)
    \item $\frac{\partial \mathbf{X}}{\partial \mathbf{R}}$: IFFT operation (linear)
    \item $\frac{\partial \mathbf{R}}{\partial s}$: Element-wise multiplication gradient
\end{itemize}

\subsection{Implementation Details}

We outline the key considerations for implementing GRAF in any automatic differentiation framework:

\begin{center}
\fbox{\begin{minipage}{0.9\textwidth}
\textbf{Algorithm 1: GRAF - Differentiable Ambiguity Function}\\[0.5em]
\textbf{Input:} Complex signal $s \in \mathbb{C}^N$\\
\textbf{Output:} Ambiguity function $\chi \in \mathbb{R}^{N \times N}$\\[0.5em]
\begin{enumerate}
\item Create shift matrix $\mathbf{S} \in \mathbb{C}^{N \times N}$:
    \begin{itemize}
    \item[] For $k = 0$ to $N-1$: $\mathbf{S}[k,:] \gets \text{circshift}(s, k)$
    \end{itemize}
\item $\mathbf{R} \gets s \odot \mathbf{S}^*$ \hfill (element-wise product with broadcast)
\item $\mathbf{X} \gets \text{FFT}(\mathbf{R}, \text{dim}=1)$ \hfill (FFT over columns)
\item $\chi \gets \text{real}(\mathbf{X} \odot \mathbf{X}^*)$ \hfill (magnitude squared)
\item $\chi \gets \text{fftshift}(\chi)$ \hfill (center zero delay/Doppler)
\item \textbf{return} $\chi$
\end{enumerate}
\end{minipage}}
\end{center}

Key implementation considerations:

\textbf{Memory efficiency}: The shift matrix construction should use advanced indexing to avoid explicit loops. For a signal of length $N$, create indices that represent circular shifts:
\begin{equation}
\mathbf{S}[k,n] = s[(n-k) \bmod N]
\end{equation}

\textbf{Numerical stability}: Normalize the output to prevent gradient explosion during backpropagation:
\begin{equation}
\chi \gets \chi / \max_{k,m} \chi[k,m]
\end{equation}

\textbf{Batch processing}: All operations should support batched inputs $s \in \mathbb{C}^{B \times N}$, where $B$ is the batch dimension (number of waveforms processed simultaneously), for efficient optimization of $B$ waveforms simultaneously.

\textbf{GPU acceleration}: The matrix operations are inherently parallelizable. Modern frameworks automatically handle device placement and kernel optimization.

\textbf{Complex gradient handling}: Ensure proper Wirtinger derivative computation for complex-valued operations. Most modern automatic differentiation frameworks (PyTorch, TensorFlow, JAX) handle this natively through their complex tensor support.

The implementation achieves $O(N^2 \log N)$ computational complexity, matching traditional methods while enabling gradient-based optimization through automatic differentiation.

\section{Loss Functions for Waveform Design}

The differentiable ambiguity function enables direct optimization of various radar performance metrics. We present several differentiable loss functions that demonstrate the versatility of GRAF for multi-objective waveform design. While these examples illustrate the framework's capabilities, our experimental validation (Section 7) focuses specifically on jointly optimizing Peak Sidelobe Level (PSL) and spectral variance for Low Probability of Intercept (LPI) applications. Researchers can combine these metrics or develop custom loss functions tailored to their specific radar requirements.

\subsection{Ambiguity Function Metrics}

We define differentiable metrics for common radar waveform requirements:

\textbf{Peak Sidelobe Level (PSL):} The PSL measures the highest sidelobe relative to the mainlobe:
\begin{equation}
\text{PSL} = \max_{(\tau,f_d) \in \Omega_s} \chi(\tau, f_d) / \chi(0, 0)
\end{equation}
where $\Omega_s$ is the sidelobe region excluding the mainlobe. While the max operation is technically non-differentiable, modern automatic differentiation frameworks like PyTorch handle this through subgradients, allowing direct optimization of the exact PSL metric without approximations.

\textbf{Integrated Sidelobe Level (ISL):} The ISL measures total sidelobe energy:
\begin{equation}
\text{ISL} = \frac{1}{\chi(0,0)} \sum_{(\tau,f_d) \in \Omega_s} \chi(\tau, f_d)
\end{equation}

This is naturally differentiable as a sum operation.

\textbf{Mainlobe Width:} We define range and Doppler resolution by the $-3$dB mainlobe width:
\begin{equation}
W_\tau = \min\{|\tau| : \chi(\tau, 0) < 0.5 \chi(0,0)\}
\end{equation}
We approximate this using a differentiable sigmoid threshold:
\begin{equation}
W_{\tau,\text{diff}} = \sum_\tau |\tau| \cdot \sigma\left(\gamma\left(\chi(\tau,0) - 0.5\chi(0,0)\right)\right)
\end{equation}
where $\sigma$ is the sigmoid function and $\gamma$ controls the transition sharpness.

\subsection{Composite Loss Function}

For practical waveform optimization, we combine multiple objectives:
\begin{equation}
\mathcal{L} = \lambda_1 \mathcal{L}_{\text{match}} + \lambda_2 \mathcal{L}_{\text{PSL}} + \lambda_3 \mathcal{L}_{\text{ISL}} + \lambda_4 \mathcal{L}_{\text{const}}
\end{equation}
where:
\begin{itemize}
    \item $\mathcal{L}_{\text{match}} = \|\chi - \chi_{\text{target}}\|_F^2$: Match target ambiguity, with $\chi_{\text{target}}$ being a designer-supplied reference ambiguity surface (e.g., the thumb-tack pattern of an ideal LFM pulse)
    \item $\mathcal{L}_{\text{PSL}} = \text{PSL}$: Minimize peak sidelobes
    \item $\mathcal{L}_{\text{ISL}} = \text{ISL}$: Minimize integrated sidelobes
    \item $\mathcal{L}_{\text{const}} = \text{var}(|s|)$: Constant modulus constraint
\end{itemize}
The weights $\lambda_i$ control the relative importance of each objective.

\section{Applications}

\subsection{Neural Network Integration}

GRAF can be integrated as a differentiable layer within neural network architectures, enabling end-to-end learning from design parameters to waveforms with desired ambiguity properties.

\begin{figure}[ht]
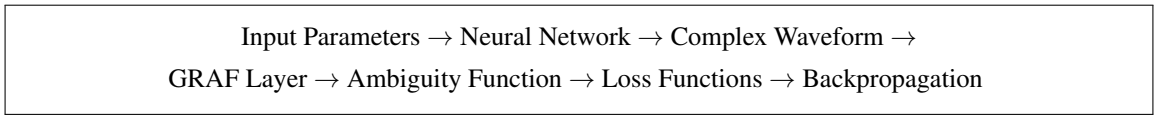

\centering
\fbox{
\begin{minipage}{0.9\textwidth}
\centering
\vspace{5pt}
Input Parameters $\rightarrow$ Neural Network $\rightarrow$ Complex Waveform $\rightarrow$ \\
\vspace{5pt}
GRAF Layer $\rightarrow$ Ambiguity Function $\rightarrow$ Loss Functions $\rightarrow$ Backpropagation
\vspace{5pt}
\end{minipage}
}
\caption{End-to-end waveform generation pipeline with GRAF}
\end{figure}

A typical architecture for waveform generation would include:
\begin{itemize}
    \item \textbf{Input layer}: Design parameters (bandwidth, duration, target characteristics)
    \item \textbf{Hidden layers}: Nonlinear transformations mapping parameters to complex waveform coefficients
    \item \textbf{Waveform layer}: Normalization to satisfy constraints (e.g., constant modulus)
    \item \textbf{GRAF layer}: Compute differentiable ambiguity function
    \item \textbf{Loss computation}: Evaluate ambiguity-based objectives
\end{itemize}

This architecture enables learning waveform generation policies that inherently satisfy ambiguity function constraints, eliminating the need for post-hoc optimization.

\subsection{Direct Waveform Optimization}

Our experimental validation (Section 7) demonstrates this approach for multi-objective optimization.

For existing waveforms, GRAF enables direct gradient-based refinement:
\begin{equation}
s^{(t+1)} = s^{(t)} - \eta \nabla_s \mathcal{L}(\chi_s)
\end{equation}

Common constraints are handled through projected gradient descent:
\begin{itemize}
    \item \textbf{Constant modulus}: Project onto the unit circle: $s \leftarrow s / |s|$
    \item \textbf{Bandwidth limits}: Apply spectral mask in frequency domain
    \item \textbf{Phase continuity}: Add smoothness regularization to the loss function
\end{itemize}

The automatic differentiation framework handles gradient computation through the GRAF operations, eliminating the need for manual derivative calculations.

\subsection{Multi-Objective Waveform Design}
While Section 5.2 showed how to combine multiple performance metrics for a single operational scenario, real-world applications often require waveforms that perform well across multiple distinct scenarios. The differentiable formulation enables simultaneous optimization for these multiple operational conditions:
\begin{equation}
\mathcal{L}_{\text{multi}} = \sum_{i=1}^{M} \lambda_i \mathcal{L}_i(\chi_s, \text{scenario}_i)
\end{equation}
where each scenario$_i$ corresponds to a distinct operational condition (e.g., target velocity, range profile), and $\lambda_i$ are designer-chosen weights controlling the importance of each scenario. Note that each $\mathcal{L}_i$ could itself be a composite loss function as defined in Equation (23).

Examples of multi-scenario optimization include:
\begin{itemize}
    \item \textbf{Multi-target environments}: Simultaneous optimization for targets at different velocities and ranges, with scenario$_i$ representing each target's range-Doppler profile
    \item \textbf{Time-varying channels}: Optimization across different propagation conditions, where each scenario$_i$ represents a different multipath or clutter environment  
    \item \textbf{Cognitive adaptation}: Real-time adjustment where scenarios represent predicted future environmental states based on sensing feedback
    \item \textbf{Robust design}: Optimization across worst-case scenarios to ensure minimum performance guarantees under uncertainty
\end{itemize}

Other applications of the GRAF framework include interference mitigation through adaptive spectral notching, MIMO radar waveform design for spatial diversity, and low probability of detection systems.

\section{Experimental Validation}

\subsection{Experimental Setup}

To validate the practical advantages of GRAF, we compare gradient-based optimization against genetic algorithms (GA)---the current standard for complex radar waveform design. We design a multi-objective optimization problem that reflects real-world radar challenges: simultaneously minimizing peak sidelobe levels while achieving low probability of intercept (LPI) characteristics.

The choice of this experimental comparison warrants explanation. Prior to our differentiable formulation, the landscape of radar waveform optimization methods was dictated by problem structure:

\textbf{Single-objective problems:} When optimizing a single metric (e.g., minimizing PSL alone), the radar community has established preferences: (1) analytical methods when closed-form solutions exist, (2) convex optimization when the problem admits convex reformulation, and (3) metaheuristic approaches like GA only when other methods fail.

\textbf{Multi-objective problems:} The situation changes dramatically for multi-objective optimization:
\begin{itemize}
    \item Analytical methods cannot handle competing objectives simultaneously
    \item Convex optimization typically fails due to non-convex radar constraints (constant modulus, spectral masks)
    \item Genetic algorithms become the \textit{de facto} standard---not by choice, but by necessity
\end{itemize}

This context explains why we specifically compare against GA for multi-objective optimization: it represents the current state-of-the-art for problems where multiple conflicting radar performance metrics must be balanced. GRAF transforms this landscape by enabling gradient-based optimization for multi-objective problems that previously required population-based methods.

All experiments were conducted on a standard laptop CPU (Intel Core i7-1165G7 @ 2.80GHz) to demonstrate practical applicability without specialized hardware. The source code for reproducing these experimental results is publicly available at \url{https://github.com/marcbara/graf-psl-lpi} and archived at \href{https://doi.org/10.5281/zenodo.15763301}{DOI: 10.5281/zenodo.15763301}.

\subsubsection{Problem Formulation}

We consider the joint optimization of detection performance and spectral uniformity:
\begin{equation}
\mathcal{L} = \text{PSL} + \lambda \cdot \text{SpectralVariance} \cdot \alpha
\end{equation}
where PSL is the peak sidelobe level ratio (linear scale), $\text{SpectralVariance} = \text{var}\{|\text{FFT}(s)|^2 / \sum|\text{FFT}(s)|^2\}$ measures the non-uniformity of the power spectrum, $\lambda$ controls the trade-off between objectives, and $\alpha = 2000$ is a scaling factor chosen such that both terms contribute equally when $\lambda = 1$.

\subsubsection{Implementation Details}

Both optimization methods operate on phase-coded waveforms with constant modulus constraint. We use $N = 256$ samples, representing realistic medium-range radar pulses at typical radar sampling rates (e.g., 10 MHz sampling yields 25.6 $\mu$s pulses suitable for 3.8 km range windows). 

The gradient-based method employs the Adam optimizer with learning rate 0.01, determined through preliminary experiments to balance convergence speed and stability. We run 2000 iterations, sufficient for convergence as demonstrated in our results (typically converging by iteration 1500). 

For GA, we use a population size of 50 individuals—a standard choice balancing diversity and computational cost—evolved over 300 generations. This yields 15,000 ambiguity function evaluations, providing GA with $7.5\times$ more function evaluations than the gradient method to ensure fair comparison despite GA's inherent inefficiency.

\subsection{Results and Analysis}

\subsubsection{Convergence Analysis}

\begin{figure}[htbp]
    \centering
    \includegraphics[width=\textwidth]{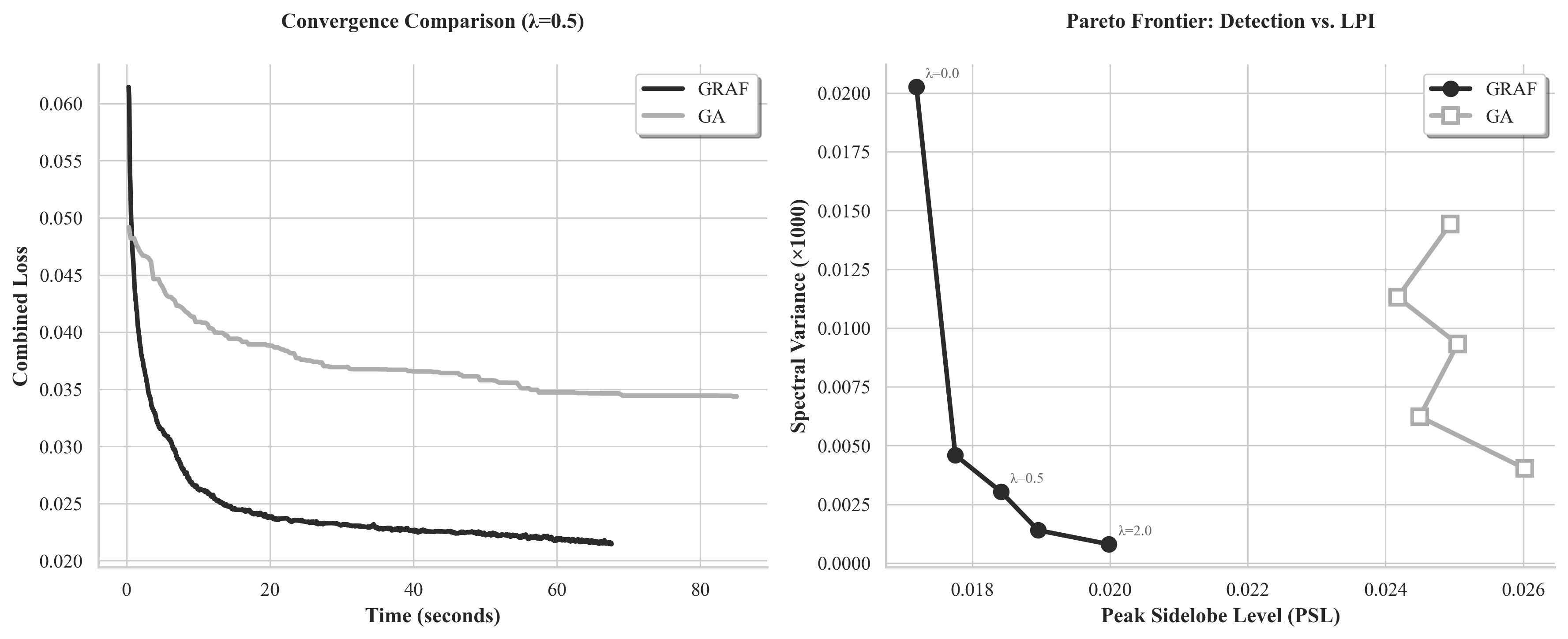}
    \caption{Convergence comparison and Pareto frontier analysis. Left: Convergence behavior for GRAF vs GA at $\lambda = 0.5$. Right: Pareto frontier showing GRAF's superior trade-offs between PSL and spectral variance across all $\lambda$ values.}
    \label{fig:fig1_main_comparison}
\end{figure}

Figure~\ref{fig:fig1_main_comparison} presents the convergence behavior for both methods at $\lambda = 0.5$. GRAF achieves rapid convergence within approximately 67 seconds, reaching a final combined loss of 0.022. In contrast, GA exhibits slow, stochastic improvement, achieving only 0.034 after 85 seconds---a $25\%$ longer computation time yet $35\%$ higher loss. The gradient method shows characteristic smooth descent, while GA's trajectory reveals the limitations of population-based search in high-dimensional spaces.

\subsubsection{Multi-Objective Optimization Performance}

The Pareto frontier analysis (Figure~\ref{fig:fig1_main_comparison}) reveals fundamental differences in optimization capability. GRAF consistently finds superior trade-offs between PSL and spectral variance across all $\lambda$ values. The GA solutions lie strictly dominated by the gradient-based frontier, indicating inferior performance on both objectives simultaneously. This dominance demonstrates that gradient-based optimization more effectively navigates the non-convex landscape of the ambiguity function.

\subsubsection{Computational Efficiency}

\begin{figure}[htbp]
    \centering
    \includegraphics[width=\textwidth]{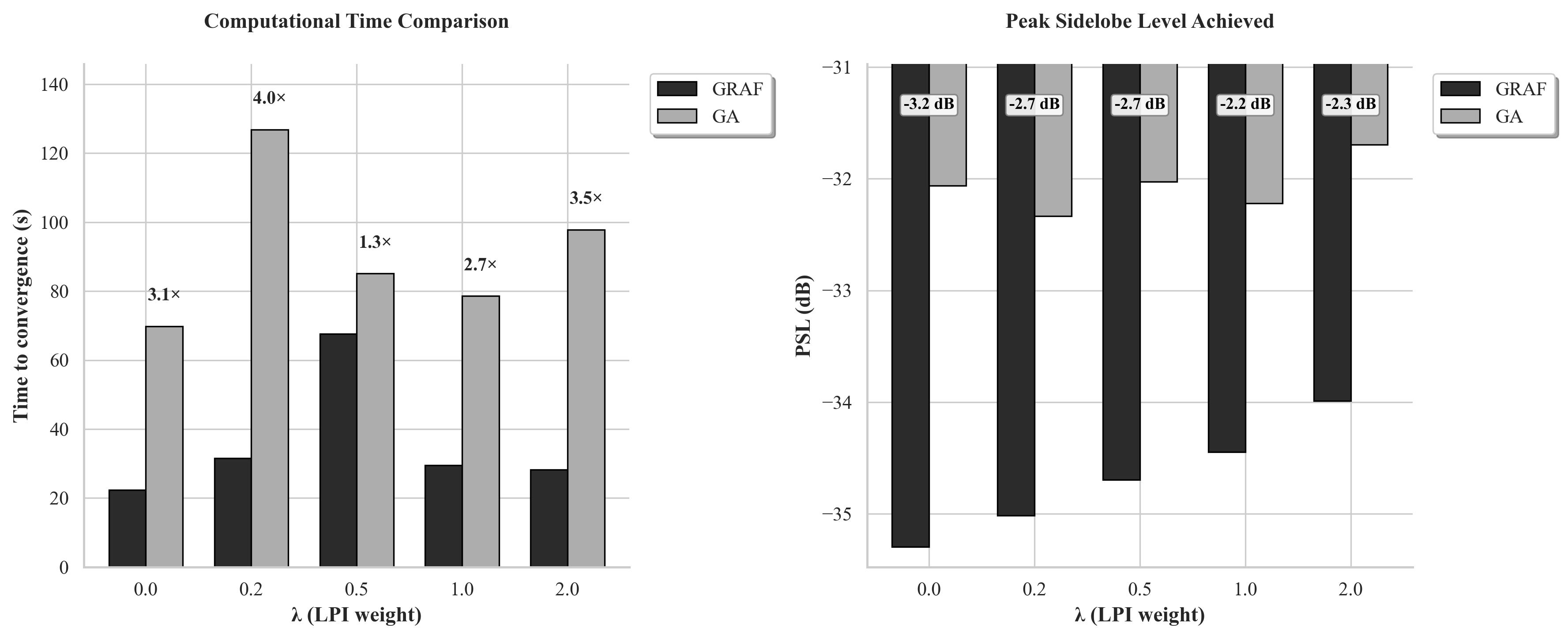}
    \caption{Computational performance comparison between GRAF and GA. Left: Time to convergence showing speedup factors. Right: Peak sidelobe level achieved with GRAF improvements indicated.}
    \label{fig:fig2_performance_summary}
\end{figure}

Figure~\ref{fig:fig2_performance_summary} summarizes the computational performance across different trade-off parameters. The results show:

\begin{itemize}
\item $\lambda = 0.0$: GRAF achieves $3.1\times$ speedup with $3.2$ dB PSL improvement
\item $\lambda = 0.25$: GRAF achieves $4.0\times$ speedup with $2.7$ dB PSL improvement  
\item $\lambda = 0.5$: GRAF achieves $1.3\times$ speedup with $2.7$ dB PSL improvement
\item $\lambda = 1.0$: GRAF achieves $2.7\times$ speedup with $2.2$ dB PSL improvement
\item $\lambda = 2.0$: GRAF achieves $3.5\times$ speedup with $2.3$ dB PSL improvement
\end{itemize}

GRAF achieves speedups ranging from $1.3\times$ to $4.0\times$ while simultaneously producing solutions that are 2.2--3.2 dB better in PSL. The speedup advantage is most pronounced for $\lambda = 0.25$, where GRAF converges $4.0\times$ faster than GA. In radar applications, a 3 dB improvement represents a $50\%$ reduction in sidelobe power, significantly enhancing target detection capability in cluttered environments.

\subsubsection{Spectral Adaptation for LPI}

\begin{figure}[htbp]
    \centering
    \includegraphics[width=\textwidth]{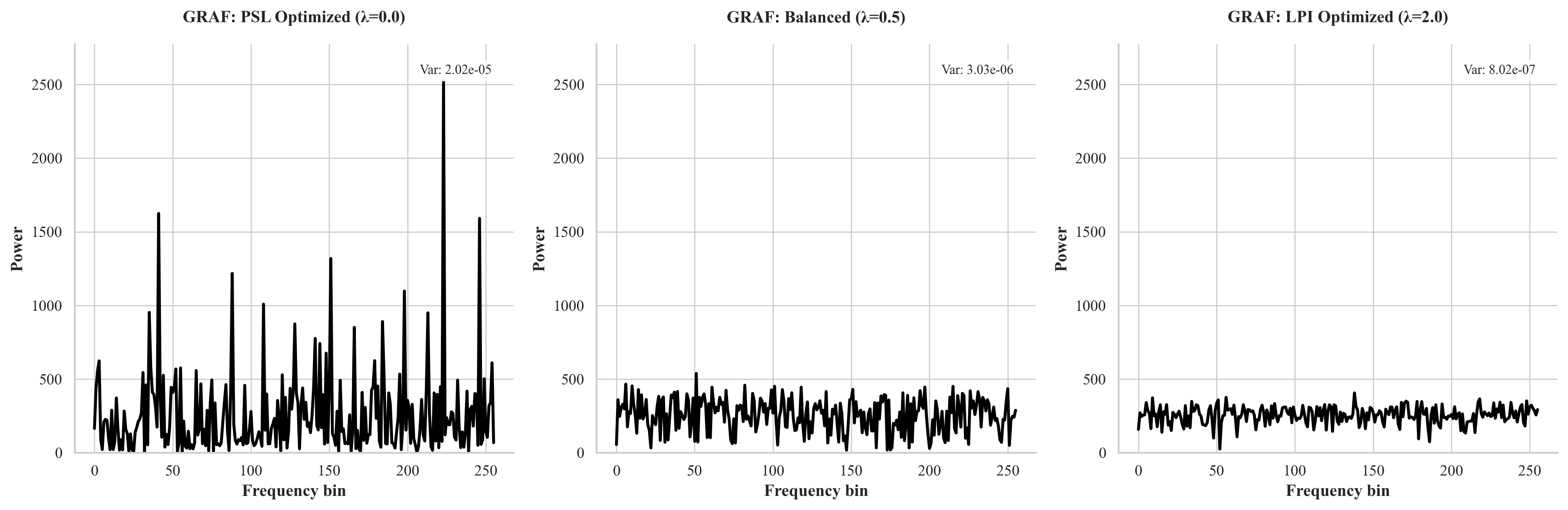}
    \caption{Spectral adaptation for LPI capability. GRAF-optimized waveforms show progressive spectral flattening as LPI weight $\lambda$ increases: (a) PSL optimized ($\lambda=0$), (b) Balanced ($\lambda=0.5$), (c) LPI optimized ($\lambda=2.0$).}
    \label{fig:fig3_spectrum_comparison}
\end{figure}

Figure~\ref{fig:fig3_spectrum_comparison} demonstrates the successful adaptation of spectral characteristics as the LPI weight $\lambda$ increases. For $\lambda = 0$ (PSL optimization only), the spectrum exhibits high concentration with distinct peaks (spectral variance $2.02 \times 10^{-5}$), making it easily detectable by electronic warfare systems. The balanced case at $\lambda = 0.5$ shows intermediate uniformity with variance $3.03 \times 10^{-6}$. As $\lambda$ increases to 2.0, the spectrum becomes progressively more uniform, with spectral variance decreasing to $8.02 \times 10^{-7}$---a 25-fold improvement over the PSL-only case. The final spectrum approaches noise-like characteristics, achieving the LPI objective while maintaining excellent radar performance.

\subsection{Discussion of Results}

The experimental validation demonstrates three key advantages of GRAF:

\textbf{Superior Convergence:} The gradient-based method consistently finds better solutions across all trade-off scenarios. Even when GA is given longer computation time, GRAF maintains a 2.2--3.2 dB PSL advantage, indicating fundamentally superior optimization capability rather than mere speed.

\textbf{Multi-Objective Capability:} The dominated Pareto frontier achieved by gradient-based optimization indicates effective balancing of competing objectives. The method successfully trades detection performance for LPI characteristics in a controlled, predictable manner while maintaining superiority in both metrics.

\textbf{Practical Feasibility:} Achieving consistent speedups on standard laptop hardware suggests even greater improvements on GPU-accelerated systems. The matrix-based operations in our implementation are inherently parallelizable, potentially enabling real-time waveform adaptation for cognitive radar applications.

\textbf{Consistent Performance:} Unlike GA's stochastic nature that produces variable results across runs, GRAF provides deterministic convergence to superior solutions, crucial for mission-critical radar applications requiring reliable performance.

These results validate that GRAF enables a new paradigm in radar waveform design, replacing slow metaheuristic searches with efficient gradient-based optimization while achieving superior performance across multiple objectives. The experimental evidence demonstrates not merely an incremental improvement, but a fundamental shift in accessible optimization techniques for practical radar waveform design. 

\section{Discussion}

\subsection{Theoretical Contributions}

Our differentiable ambiguity function formulation addresses fundamental computational challenges that have prevented the integration of classical radar theory with modern machine learning frameworks. The approach provides the first rigorous treatment of Wirtinger derivatives for radar ambiguity function computation, enabling proper gradient flow through complex-valued operations while maintaining mathematical equivalence to traditional formulations.

The conceptual simplicity of our solution—applying FFT operations over the time dimension with proper gradient handling—belies the technical barriers that have prevented its implementation. The intersection of complex-valued signal processing, Wirtinger calculus, and automatic differentiation represents a knowledge gap between communities. While radar engineers possess deep understanding of ambiguity functions, they typically work outside modern ML frameworks. Conversely, ML practitioners rarely encounter the complex-valued operations and circular convolution structures fundamental to radar signal processing. Our contribution bridges this divide with a practical implementation that makes each community's tools accessible to the other.

The reformulation using parallelized FFT operations achieves computational complexity suitable for practical applications. The efficient indexing strategies enable GPU acceleration while maintaining numerical stability throughout the gradient computation pipeline. This represents a significant advance over existing approaches that rely on explicit delay loops or finite-difference approximations.

Unlike existing gradient-based approaches that derive analytical gradients for specific optimization problems, our implementation provides a general-purpose differentiable component that can be composed with arbitrary differentiable operations. This modularity enables researchers to incorporate ambiguity function constraints into diverse machine learning architectures without requiring custom gradient derivations for each application.

\subsection{Enabling New Research Directions}

GRAF opens several previously inaccessible research avenues. Neural networks can now incorporate ambiguity function constraints directly into their loss functions, enabling joint optimization of waveform parameters and ambiguity characteristics. This capability supports the development of learning-based waveform libraries adapted to specific operational environments.

Real-time waveform adaptation based on environmental feedback becomes computationally feasible through gradient-based optimization. The efficient forward and backward passes enable cognitive radar systems to adapt waveforms within practical timing constraints, supporting Haykin's vision of truly adaptive radar systems~\cite{haykin2006cognitive}. The differentiable formulation naturally handles conflicting objectives through weighted loss functions, enabling simultaneous optimization of range resolution, Doppler tolerance, sidelobe levels, and spectral constraints.

The mathematical framework extends beyond traditional radar to applications including joint radar-communication systems, MIMO radar optimization~\cite{chen2025ambiguity,qiu2025constrained}, and integration with other sensing modalities. Each of these domains can leverage the established infrastructure for automatic differentiation rather than developing custom optimization approaches.

\subsection{Implementation Considerations}

Several practical considerations emerge from our implementation approach. The O(N²) memory requirement for the circulant matrix construction limits maximum waveform length to approximately N $\approx$ 4096 samples on current GPU hardware. Future work could explore block-based processing or sparse representations for longer sequences.

Our implementation using float32 or complex64 provides adequate numerical stability for most applications. The normalization strategies prevent gradient explosion while maintaining sufficient dynamic range for optimization. For operations requiring differentiable approximations (such as the mainlobe width computation shown in Section 5.1), sigmoid-based formulations provide controlled trade-offs between approximation accuracy and gradient quality.

Forward pass computation scales favorably with waveform length, enabling real-time applications. The automatic differentiation overhead is modest compared to the computational benefits of gradient-based optimization over population-based methods, making the approach practical for iterative design workflows.

While our experimental validation focuses on PSL and spectral variance optimization as a concrete demonstration, the GRAF framework's modular design supports arbitrary differentiable loss functions. Researchers can leverage this flexibility to address diverse radar waveform design challenges, from MIMO systems to joint radar-communication applications, by simply defining appropriate differentiable objectives within the framework.

\subsection{Comparison with Existing Approaches}

Our approach differs fundamentally from existing gradient-based methods in the radar literature. While methods like Alhujaili's quartic gradient descent~\cite{alhujaili2020quartic} derive analytical gradients for specific cost functions, our implementation provides a general-purpose differentiable ambiguity function that can be combined with arbitrary differentiable operations. This generality comes without sacrificing computational efficiency or mathematical rigor.

Existing approaches require custom optimization code and manual gradient derivation for each new problem formulation~\cite{mohr2021gradient,xiao2025unimodular}. Our implementation enables seamless integration with modern machine learning pipelines, reducing the barrier to entry for researchers seeking to apply gradient-based methods to radar waveform design problems.

The modular design philosophy supports rapid experimentation and prototyping, enabling researchers to focus on algorithm development rather than gradient computation details. This represents a significant practical advantage over traditional approaches that require substantial implementation effort for each new optimization variant.

\subsection{Experimental Insights}

Our experimental validation provides concrete evidence for GRAF's practical advantages. The consistent speedup factors across different optimization scenarios demonstrate that the theoretical computational efficiency translates to real-world performance gains. More significantly, the simultaneous improvement in solution quality indicates that gradient-based optimization not only converges faster but also more effectively navigates the non-convex landscape of the ambiguity function.

The successful optimization of competing objectives (PSL vs. spectral variance) validates GRAF's ability to handle multi-objective problems that have traditionally required population-based methods. This opens new possibilities for real-time adaptive radar systems that must balance multiple performance criteria.

\subsection{Limitations and Future Work}

The circular convolution formulation assumes periodic waveforms, which may not be appropriate for all applications. Extension to non-periodic waveforms requires alternative formulations that maintain computational efficiency while handling boundary conditions appropriately.

The discrete implementation samples the continuous ambiguity function, potentially missing features between sample points. Adaptive sampling strategies could improve accuracy for critical applications, though this would require careful consideration of computational trade-offs.

Extension to cross-ambiguity functions for MIMO radar applications represents an important future direction~\cite{niu2025novel}. The mathematical framework provides a foundation for such extensions, though implementation details require careful consideration of memory scaling and computational complexity.

While the implementation is suitable for GPU acceleration, deployment on specialized radar hardware requires consideration of fixed-point arithmetic and memory constraints. Hardware-specific optimizations could enable real-time waveform adaptation on embedded radar platforms.

As the first differentiable ambiguity function implementation, GRAF opens entirely new research directions. Future work by the community can now explore applications made possible by this framework: neural network integration for learned waveform generation, real-time cognitive radar adaptation, and seamless incorporation into end-to-end radar system optimization. The modular, open-source nature of GRAF enables researchers to extend our framework to specialized applications such as MIMO radar, joint radar-communication systems, and adaptive interference mitigation—domains previously inaccessible to gradient-based methods.

\section{Conclusion}

We have presented GRAF (Gradient-based Radar Ambiguity Functions), the first complete mathematical framework and computational implementation for differentiable radar ambiguity functions. By addressing the fundamental technical challenges of complex differentiation, computational efficiency, and numerical stability, this work enables the integration of classical radar theory with modern machine learning frameworks.

The approach provides a mathematically rigorous differentiable formulation that maintains complex gradient flow throughout the computation pipeline. By applying FFT operations over the time dimension for each delay, we achieve exact computation of the ambiguity function while maintaining differentiability. The efficient implementation using parallelized FFT operations achieves computational complexity suitable for practical applications. The integration methodology enables seamless use within neural network architectures and automatic differentiation frameworks, while comprehensive treatment of differentiable metrics supports common radar waveform design objectives.

The resulting implementation provides a foundational tool for next-generation radar systems that can learn and adapt their waveforms based on environmental feedback and operational requirements. By enabling gradient-based optimization of ambiguity function characteristics, this work opens the door to applying modern machine learning techniques to radar waveform design.

This research establishes the mathematical and computational foundation for new directions in cognitive radar, including learning-based waveform libraries, real-time environmental adaptation, joint radar-communication waveform design~\cite{wang2022joint}, and MIMO radar optimization. 

GRAF represents a bridge between classical radar signal processing and modern machine learning, enabling researchers to leverage decades of radar theory within contemporary optimization frameworks. As cognitive radar systems evolve toward adaptive, learning-based architectures, this foundational tool provides the mathematical and computational infrastructure necessary for continued innovation at the intersection of radar systems and artificial intelligence.

\bibliographystyle{plain}
\bibliography{references}

\end{document}